Bound State Solutions of Schrödinger Equation for a more general Woods-Saxon Potential with Arbitrary $l$ - state.

Akpan N.Ikot*[1] and Ita O.Akpan[2]

[1] Theoretical Physics Group, Department of Physics, University of Uyo, Nigeria

[2] Department of Physics, University of Calabar, Nigeria

E-mail: ndemikot2005@yahoo.com

Abstract

The energy spectra and the wave function depending on the c-factor are investigated for a more general Woods-Saxon potential (MGWSP) with an arbitrary $l$ - state. The wave functions are expressed in terms of the Jacobi polynomials. Two potentials are obtained from this MGWSP as special cases. These special potentials are Hulthen and the standard Woods-Saxon potentials. We also discuss the energy spectrum and wave function for the special cases.

1. **Introduction**

In recent times different methods have been adopted for the solution of Schrödinger equation (SE) with various potentials [1-5] such methods include numerical and analytical techniques [6-7], suppersymmetry (SUSSY) [8], Pekevis approximation [9] and the Nikiforov-Uvarov method [10]. An exact solution of the SE is of high importance in non-relativistic quantum mechanics.

However, there are very few potentials for which the radial SE can be solved exactly for all n and $l$ – values. Levai and Williams in their paper developed a simple method for constructing potential for which the SE can be solved exactly in terms of special functions [11]. The exact solutions of SE for the Woods-Saxon potential(WSP)[12] cannot be solved exactly for $l \neq 0$, though Flugge [13] gave an exact expression for the wave function but uses graphical method to obtain the energy eigen values for $l = 0$ [14].

The WSP is one of the short-range potentials well-known in Physics and plays a vital role in nuclear and particle Physics, atomic Physics, condensed matter and chemical Physics [15].

In this paper, we attempt to solve the radial SE for this more general Woods-Saxon potential (MGWSP) using the Nikiforov-Uvarov method and obtain the wave function and the corresponding eigenvalues for arbitrary $l$ – states and then deduce some well-known potentials from our results.

**2.     Review of Nikiforov-Uvarov method**

The standard form of the Nikiforov-Uvarov method equation is written as [10]

$$\psi''(s) + \frac{\bar{\tau}(s)}{\sigma(s)}\psi'(s) + \frac{\bar{\sigma}(s)}{\sigma^2(s)}\psi(s) = 0 \qquad (1)$$

where $\sigma(s)$ and $\bar{\sigma}(s)$ are polynomial, of the second degree at most and $\bar{\tau}(s)$ is a first degree polynomials. Thus, from equation (1), the Schrödinger equation and the Schrodinger like equations can be solved analytically by means of the special potentials with this method. Hence, in order to obtain the particular solution of equation (1), one can use the transformation for the wave function as

$$\psi(s) = \varphi(s)\chi_n(s) \qquad (2)$$

This reduces equation (1) to an equation of hypergeometric type as

$$\sigma(s)\chi_n''(s) + \bar{\tau}(s)\chi_n'(s) + \lambda\chi_n(s) = 0 \qquad (3)$$

and $\varphi(s)$ is defined as a logarithmic derivative in the form

$$\frac{\varphi'}{\varphi(s)} = \frac{\pi(s)}{\sigma(s)} \qquad (4)$$

The function $\pi(s)$ and the parameter $\lambda(s)$ required for the Nikiforov-Uvarov method

$$\pi(s) = \frac{\sigma'-\bar{\tau}(s)}{2} \pm \sqrt{\left(\frac{\sigma'-\bar{\tau}}{2}\right)^2 - \bar{\sigma}(s) + k\sigma(s)} \qquad (5)$$

$$\lambda(s) = k + \pi'(s) \qquad (6)$$

Conversely, in order to find the values of k in equation (5), then the expression under the square root most be the square of the polynomials.

Consequently, the eigenvalue equation for the Schrödinger equation becomes,

$$\lambda_n = -n\tau'(s) - \frac{n(n-1)}{2}\sigma''(s) \qquad (7)$$

where

$$\tau(s) = \bar{\tau}(s) + 2\pi(s) \qquad (8)$$

and its derivative is negative. The other wave function can be determine using the Rodriques Relation

**3. Bound State solution of the Radial Schrödinger Equation with MGWSP.**

The standard Wood-Saxon potential is defined as [16]

$$V(r) = \frac{V_0}{\left[1+exp\left(\frac{r-R_0}{a}\right)\right]}, a \ll R_0 \qquad (9)$$

The Schrödinger equation for the potential V(r) is of the form [16]

$$\left(\frac{-\hbar^2}{2m}\nabla^2 + V(r)\right)\psi(r) = E\psi(r), \tag{10}$$

where $\psi(r)$ is the wave function, E is the energy eigenvalues, $\hbar$ is the Planck Constant, m is the mass and $\nabla^2$ is the Laplacian operator. The radial Schrödinger equation [17] of equation (10) with Woods-Saxon potential is given by

$$\frac{d^2\psi(r)}{dr^2} + \frac{2m}{\hbar^2}\left[E + \frac{V_0}{[1+\exp(2\alpha(r-R_0))]}\right]\psi(r) + \frac{l(l+1)}{r^2}\psi(r) = 0 \tag{11}$$

where $l$ is the angular momentum quantum number and $\alpha = 1/2a$. Writing the wave function $\psi(r) = R(r)/r$ reduces equation (11) into the form

$$\frac{d^2R(r)}{dr^2} + \left[E + \frac{V_0}{(1+e^{2\alpha(r-R_0)})} - \frac{l(l+1)\hbar^2}{2\mu r^2}\right]R(r) = 0 \tag{12}$$

Equation (12) can be expressed in terms of the effective potential as

$$\frac{d^2R(r)}{dr^2} + \frac{2m}{\hbar^2}[E - V_{eff}(r)]R(r) = 0 \tag{13}$$

where the effective potential $V_{eff}(r)$ is defined as

$$V_{eff}(r) = V(r) + \frac{l(l+1)\hbar^2}{2\mu r^2} \tag{14}$$

where the second term corresponds to the centrifugal term.

Following the new improved approximation scheme [18] for the centrifugal term, we expressed the effective potential as

$$V_{eff}(r) = V(r) + \frac{2l(l+1)\hbar^2}{\mu\alpha^2}\frac{e^{-2\alpha r}}{(1-e^{-2\alpha r})^2} \tag{15}$$

Taking a new co-ordinate

$$s = c(\exp(2\alpha r - 1)) \tag{16}$$

where $c = \exp(-2\alpha R_0)$ and it is the c-factor that characterized the behavior of the potential as will be seen latter. Equation (16) transforms equation (13) into the form

$$(c+s)^2 \frac{d^2R}{ds^2} + (c+s)\frac{dR}{ds} + \frac{4m}{\hbar^2\alpha^2}\left[E + \frac{V_0}{(c+1+s)} - \delta_0 - \frac{\delta_0(c+s)^2}{s^2}\right]R(s) = 0, \quad (17)$$

where $\delta_0 = \frac{2\hbar^2 l(l+1)}{\mu\alpha^2}$.

Simplifying equation (17) yields

$$\frac{d^2R}{ds^2} + \frac{s}{s(c+s)}\frac{dR}{ds} + \frac{1}{s^2(c+s)^2}[(\varepsilon^2 + \gamma^2 + \xi_1^2 - \xi_4^2)s^2$$

$$+ (\beta^2 + \xi_3^2)s + \xi_2^2]R(s) = 0 \quad (18)$$

where the following dimensionless parameters have been used in obtaining equation (18);

$$\varepsilon^2 = \frac{-mE}{2\hbar^2\alpha^2}, \quad \beta^2 = \frac{mV_0}{2\hbar^2\alpha^2(1+c)}, \quad \gamma^2 = \frac{mV_0}{2\alpha^2\hbar^2(1+c)^2}, \quad \xi_1^2 = \frac{m\delta_0}{2\hbar^2\alpha^2},$$

$$\xi_2^2 = \frac{mc^2\delta_0}{2\hbar^2\alpha^2}, \quad \xi_3^2 = \frac{mc\delta_0}{\hbar^2\alpha^2}, \quad \xi_4^2 = \frac{m}{2\hbar^2\alpha^2}. \quad (19)$$

It is pertinent at this point to note that in obtaining equation (18), we have Taylor expanded the term $(c+1+s)^{-1}$ up to second order, using the expression [17]

$$\frac{1}{c+1+s} = \frac{1}{(c+1)}\sum_{k=1}^{\infty}\frac{(-1)^k s^k}{(c+1)^k} \quad (20)$$

Now comparing equation (18) with equation (1), we obtain the following polynomials:

$$\bar{\tau} = s, \quad \sigma(s) = s(c+s), \quad \bar{\sigma}(s) = -as^2 + bs + d, \quad (21)$$

where

$$a = (\varepsilon^2 + \gamma^2 + \xi_1^2 + \xi_4^2),$$

$$b = (\beta^2 + \xi_3^2)$$

$$c = \xi_2^2 \quad (22)$$

In the NU-method, the new $\pi(s)$ is defined as

$$\pi(s) = \frac{c+s}{2} \pm \frac{1}{2}\sqrt{(4k-b_1)s^2 + (b_2 + 4kc)s + b_3} \quad (23)$$

where $b_1 = -4a + 1$, $b_2 = 2c - 4b$ and $b_3 = c^2 - 4d$.

The discriminant in the expression under the square root has to be zero. Thus, the expression becomes the square of a polynomial of first degree,

$$16c^2 k^2 + (8b_2 c - 16 b_2)k + 4\eta_1 b_2 + \eta_2^2 = 0 \tag{24}$$

When we imposed the basic requirement with respect to the constant k, its double are obtained as

$$k_\pm = \frac{b_2(2-c)}{4c^2} \pm \frac{1}{2c^2}\sqrt{b_2\big((1-c)b_2 + b_1 c^2\big)} \tag{25}$$

Substituting $k_\mp$ into Eq. (23), the following possible solutions are obtained for $\pi(s)$

$$\pi(s) = \frac{c+s}{2} \pm \frac{1}{2c^2} \begin{cases} \sqrt{\eta_2} s - \sqrt{(1-c)\eta_2 + b_1 c^2}, & \text{for } k = \frac{b_2(2-c)}{4c^2} - \frac{1}{2c^2}\sqrt{b_2(1-c)\eta_2 + b_1 c^2} \\ \sqrt{\eta_2} s - \sqrt{(1-c)\eta_2 + b_1 c^2}, & \text{for } k = \frac{b_2(2-c)}{4c^2} + \frac{1}{2c^2}\sqrt{b_2(1-c)\eta_2 + b_1 c^2} \end{cases} \tag{26}$$

For the polynomial of $\tau = \bar{\tau} + 2\pi$ which has a negative derivative, we select,

$$k(s) = \frac{b_2(2-c)}{4c^2} \pm \frac{1}{2c^2}\sqrt{b_2\big((1-c)b_2 + b_1 c^2\big)} \tag{27}$$

$$\pi(s) = \frac{c+2}{2} - \frac{1}{2c^2}\left(\sqrt{\eta_2} s - \sqrt{(1-c)\eta_2 + b_1 c^2}\right) \tag{28}$$

Therefore, with this selection and using $\lambda = k + \pi'$, we obtain the $\tau$ and $\lambda$ values as

$$\tau(s) = \left(\frac{-\sqrt{b_2}}{c^2} + c + 2\right)s + \frac{2}{c^2}\sqrt{(1-c)b_2 + b_1 c^2} \tag{29}$$

$$\lambda = \frac{\eta_2(2-c)}{4c^2} - \frac{1}{2c^2}\sqrt{\eta_2(1-c)b_2 + b_1 c^2} + \frac{c}{2} - \frac{b_2}{2c^2} \tag{30}$$

Now using Eq. (7), we have

$$\lambda_n = \frac{n\sqrt{\eta_2}}{c^2} - c - 2 - n(n-1) \tag{31}$$

Comparing Eqs. (30) and (31), we obtain the exact energy eigenvalues as

$$E_n = 1 + \delta_0 + \frac{V_0}{(1+c)^2} - \frac{1}{2}\left(\frac{\hbar^2 \alpha^2}{m}\right) - \frac{1}{4}\left(\frac{2\hbar\alpha}{m}\right)^2 \left[2c - \frac{4m}{\hbar^2 \alpha^2}\left(\frac{V_0}{1+c} + c\delta_0\right)\right](c-1)$$

$$-\frac{2\hbar^2\alpha^2}{m}\left[\frac{2n}{n\sqrt{2c-\frac{2m}{\hbar^2\alpha^2}\left(\frac{V_0}{1+c}+c\delta_0\right)}}\right.$$

$$\left.+\frac{4+c-3n(n-1)}{2\left[2c-\frac{2m}{\hbar^2\alpha^2}\left(\frac{V_0}{1+c}+c\delta_0\right)\right]}-\frac{1}{c}\right]^2 ,\text{for } c \gg 1 \qquad (32)$$

Equation (32) is the energy spectrum for the Schrödinger equation with the MGWSP. However, Eq. (32) can be rewritten for a case $c \ll 1$ as,

$$E_n = 1 + \delta_0 - \frac{1}{2}\left(\frac{\hbar^2\alpha^2}{m}\right) + V_0(1-2c)$$

$$+\frac{1}{4m}(\hbar\alpha)^2\left[2c - \frac{2m}{\hbar^2\alpha^2}(V_0(1-c) + 2\delta_0 c)\right](c-1)$$

$$-\frac{2\hbar^2\alpha^2}{m}\left[\frac{2n}{c\sqrt{2c-\frac{2m}{\hbar^2\alpha^2}(V_0(1-c)+c\delta_0)}}\right.$$

$$\left.+\frac{4+c-3n(n-1)}{2\left[2c-\frac{2m}{\hbar^2\alpha^2}(V_0(1-c)+c\delta_0)\right]}-\frac{1}{c}\right] \qquad (33)$$

In other to find the wave function we first evaluate for the weight function from Eq. (18) as

$$\rho(s) = s^{(c-\nu)}(c+s)^{\left(\frac{\nu-\mu c}{c^2}\right)} \qquad (34)$$

Where $\mu = \left(-\frac{\sqrt{\eta_2}}{c^2} + c + 2\right)$ and $\nu = \frac{2}{c^2}\sqrt{(1-c)\eta_2 + \eta_1 + c^2}$

Now using the Rodrique relation of Eq. (18b)

$$\chi_n(s) = N_n s^{-\left(\frac{c-\nu}{c}\right)}(c+s)^{\left(\frac{\nu-\mu c}{c^2}\right)}\frac{d^n}{ds^n}\left[s^{\left(n+\frac{c-\nu}{c}\right)}(c+s)^{\left(n+\frac{\nu-\mu c}{c^2}\right)}\right] \qquad (35)$$

where $N_n$ is the normalization constant. The other wave function is obtain from Eq. (4) as

$$\varphi(s) = s^{\left(\frac{c-2\nu}{2c}\right)}(c+s)^{\left(\frac{\sqrt{\eta_2}+2\nu}{2c^2}\right)} \qquad (36)$$

Therefore, radial wave function of the Schrödinger equation with the MGWSP can be written as

$$R(s) = \varphi(s)\chi_n$$

$$= C_n s^{A/2}(c+s)^{\frac{\mu}{c}+B-\sqrt{\eta_2}} \frac{d^n}{ds^n}\left[s^{n+A/2}(c+s)^{n+B-\mu c}\right] \qquad (37)$$

where $A = 2\mu/c - 1$ and $B = 2\nu/c^2$. The Wave function can be express in terms of Jacobi polynomials as

$$R(s) = C_n s^{A/2}(c+s)^{B-\sqrt{\eta_2}} P_n^{(A/2, B)}(s) \qquad (38)$$

where $N_n$ is the new normalization constant and obey the normalization condition, $\int_{-1}^{1}(R(s))^2 ds = 1$.

Finally, we write the total radial wave function as

$$R(r) = \frac{1}{r}[c(\exp(2\alpha r) - 1)]^{A/2}[c(\exp(2\alpha r))]^{B-\sqrt{\eta_2}} P_n^{(A/2, B)}(r). \qquad (39)$$

**4.  Results and Discussion**

In the Schrödinger equation with a MGWSP, we obtained two energy spectra ($c \ll 1$ and $c \geq 1$), and unnormalized wave function expressed in terms of Jacobi polynomials. Two special potentials are deduced:

**4.1 Hulthen Potential.**

The Hulthen potential [19] is obtained from the MGWSP by setting $c = -1$ and the corresponding energy eigenvalues and the wavefunction are obtained from Eq.(33) and Eq.(39) as

$$E_n = 1 + \delta_0 - \frac{1}{2}\left(\frac{\hbar^2 \alpha^2}{m}\right) + 3V_0$$

$$-\frac{1}{2m}(\hbar\alpha)^2\left[-2 - \frac{4m}{\hbar^2\alpha^2}(V_0 - \delta_0)\right]$$

$$+\frac{2\hbar^2\alpha^2}{m}\left[\frac{2n}{\sqrt{-2-\frac{2m}{\hbar^2\alpha^2}(2V_0-\delta_0)}}\right.$$

$$+ \frac{3(1-n(n-1))}{2\left[-2-\frac{2m}{\hbar^2\alpha^2}(2V_0-\delta_0)\right]} + 1 \Bigg] \quad , \tag{40}$$

and

$$R(r) = \frac{1}{r}\left[1 + (exp(2\alpha r))\right]^{A/2}\left[(\exp(2\alpha r))\right]^{B-\sqrt{\eta_2}} P_n^{(A/2,B)}(r) , \tag{41}$$

respectively.

### 4.2 Standard Woods-Saxon Potential.

For $c = 1$ the MGWSP changes to the standard WS potential [20]. The corresponding energy eigenvalues and wavefunction for this potential are obtain from equations (32) and (39) as

$$E_n = 1 + \delta_0 + \frac{V_0}{4} - \frac{1}{2}\left(\frac{\hbar^2\alpha^2}{m}\right)$$

$$- \frac{2\hbar^2\alpha^2}{m}\left[\frac{2n}{n\sqrt{2-\frac{2m}{\hbar^2\alpha^2}\left(\frac{V_0}{2}+\delta_0\right)}} + \frac{5-3n(n-1)}{2\left[2-\frac{2m}{\hbar^2\alpha^2}\left(\frac{V_0}{2}+\delta_0\right)\right]} - 1\right]^2, \tag{42}$$

and

$$R(r) = \frac{1}{r}\left[(exp(2\alpha r) - 1)\right]^{A/2}\left[(\exp(2\alpha r))\right]^{B-\sqrt{\eta_2}} P_n^{(A/2,B)}(r) , \tag{43}$$

respectively.

### 5. Conclusions

Using NU method we have discussed the approximate solution of the Schrodinger equation for a MGWSP for arbitrary $l$−states. We have obtained the energy eigenvalues equation and wavefunction of the Schrodinger for the MGWSP for $c \ll 1 \: and \: c \gg 1$. In addition, as a special case Hulthen potential and standard WS potential were discussed. Finally, it can be seen that these results are consistent with those in the literature [21].

### Acknowledgement.

This research work was partially supported by the Nandy-Rangers grant no.347-094